\documentclass[twocolumn,aps,floatfix,superscriptaddress]{revtex4}
\usepackage{amsmath,amssymb,eucal,graphicx,float}
\begin{document}
\title{Escape and Finite-Size Scaling in Diffusion-Controlled Annihilation}
\author{E.~Ben-Naim}
\affiliation{Theoretical Division and Center for Nonlinear Studies,
Los Alamos National Laboratory, Los Alamos, New Mexico 87545, USA}
\author{P.~L.~Krapivsky}
\affiliation{Department of Physics, Boston University, Boston,
  Massachusetts 02215, USA}
\begin{abstract}
We study diffusion-controlled single-species annihilation with a
finite number of particles.  In this reaction-diffusion process, each
particle undergoes ordinary diffusion, and when two particles meet,
they annihilate.  We focus on spatial dimensions $d>2$ where a finite
number of particles typically survive the annihilation process.  Using
the rate equation approach and scaling techniques we investigate the
average number of surviving particles, $M$, as a function of the
initial number of particles, $N$. In three dimensions, for instance,
we find the scaling law $M\sim N^{1/3}$ in the asymptotic regime 
$N\gg 1$. We show that two time scales govern the reaction kinetics: the
diffusion time scale, $T\sim N^{2/3}$, and the escape time scale,
$\tau\sim N^{4/3}$. The vast majority of annihilation events occur on
the diffusion time scale, while no annihilation events occur beyond
the escape time scale.
\end{abstract}
\maketitle

Reaction-diffusion processes are found in most areas of science
including biology, chemistry, physics, and geophysics
\cite{pg,js,skf,pk,bh}. Typically, in diffusion-controlled reactions,
particles diffuse in space and a reaction occurs when two or more
particles ``meet'' \cite{otb,mvs,rz,sc}. Being strongly interacting
many-body systems, reaction-diffusion processes play a central role in
pattern formation \cite{cg,vvs} and in non-equilibrium statistical
mechanics \cite{thv,krb}.

Studies of simplified reaction schemes such as annihilation,
coalescence, and aggregation show that there are two types of
behavior. In sufficiently low spatial dimensions, significant spatial
fluctuations develop and slow down the reaction kinetics
\cite{bh,otb,krb,zo,bg,pgg,tm,tw,zr,kr,jls,brt,gleb89,bbd}.  This
effect has been confirmed, both qualitatively and quantitatively, in a
number of experiments \cite{kfs,addlwsylb}.  In large spatial
dimensions spatial fluctuations are minor, and the standard rate
equation approach is applicable.  The critical dimension which
differentiates these two regimes of behavior depends on the reaction
scheme \cite{krb}.

The vast literature on non-equilibrium statistical mechanics of
reacting systems is focused on spatially-homogeneous systems where the
number of particles is infinite. Moreover, most of the theoretical
methods used to describe reaction processes apply to
spatially-homogeneous systems \cite{tm,zr,jls,brt,bbd}. While there
are a few exceptional studies of spatially-inhomogeneous
\cite{lps,kb}, or finite \cite{kpwh} systems, little attention has
been given to systems with a finite number of particles.

Here, we investigate diffusion-controlled reaction processes with a
finite, yet large, number of particles in an unbounded space. We focus
on the simplest possible reaction, single-species annihilation which
is represented by the reaction scheme
\begin{equation}
\label{process}
A+A\to \emptyset\,.  
\end{equation}

\begin{figure}[t]
\vspace{.1in}
\includegraphics[width=0.34\textwidth]{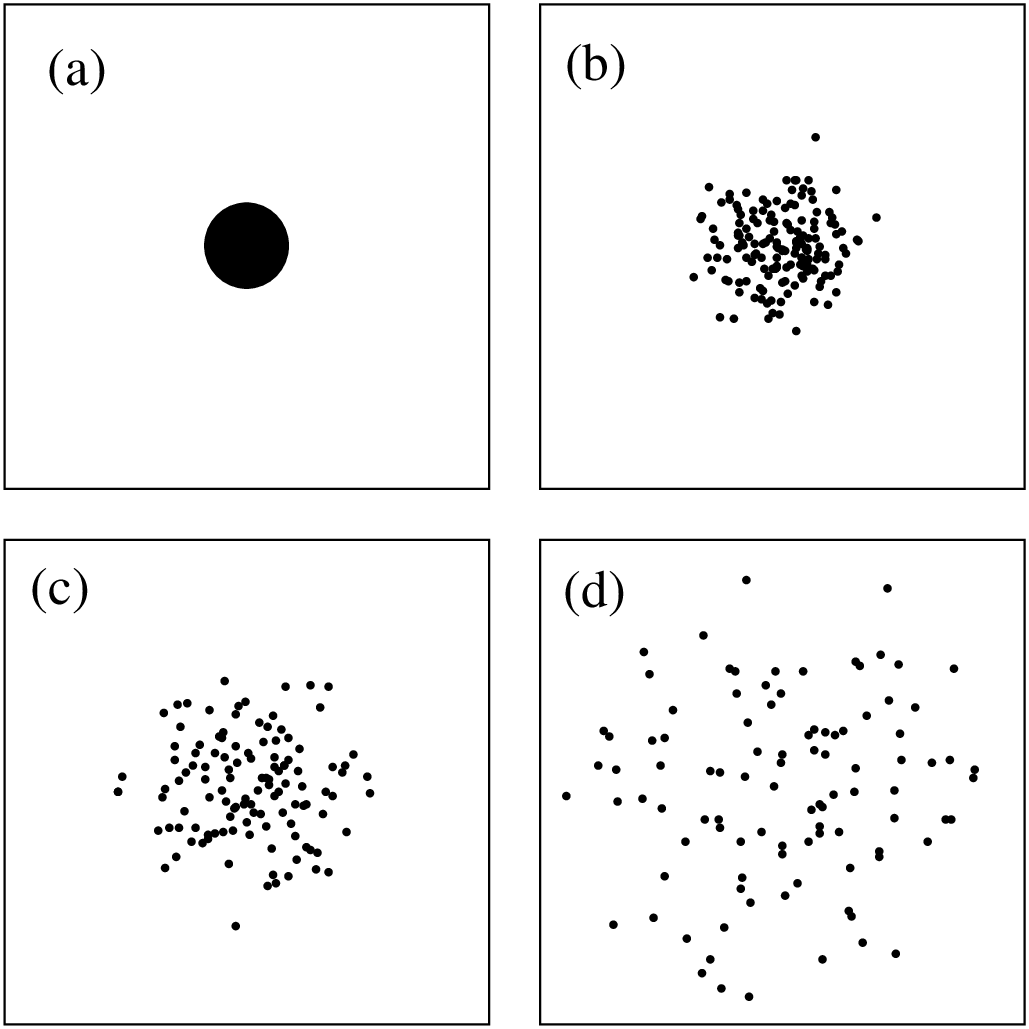}
\caption{Illustration of the diffusive escape process: (a) 
  the initial condition with $N=38,911$ particles inside a ball
  of radius $R=21$ in three dimensions; (b), (c), \& (d) show
  particle positions when $4.1\%$, $3.1\%$, and $2.5\%$ of the
  initial particles remain.  Particle positions are projected from
  three dimensions onto two dimensions.}
\label{fig-ill}
\end{figure}

In the annihilation process \eqref{process}, identical particles,
denoted by $A$, undergo Brownian motion and whenever two particles
come into contact, they disappear. We focus on the initial condition
where $N$ particles occupy a $d-$dimensional ball (see
Fig.~\ref{fig-ill}).  We also briefly discuss initial conditions where
the occupied region has an intrinsic dimension smaller than
$d$.

We are especially interested in the ultimate fate of the system, that
is, the behavior in the limit $t\to\infty$. This behavior follows from
the recurrence properties of a single random walk \cite{krb,mp}.  A
random walk on an ordinary lattice in dimension $d$ is guaranteed to
return to its starting site if and only if $d\leq 2$.  Since the
separation between a pair of particles also undergoes a random walk,
the two particles are guaranteed to meet if and only if $d\leq 2$.
Therefore in single-species annihilation, all particles disappear when
$d\leq 2$, while a finite number {\it eternally} survive when $d>2$
\cite{odd}.

\begin{figure}[t]
\includegraphics[width=0.5\textwidth]{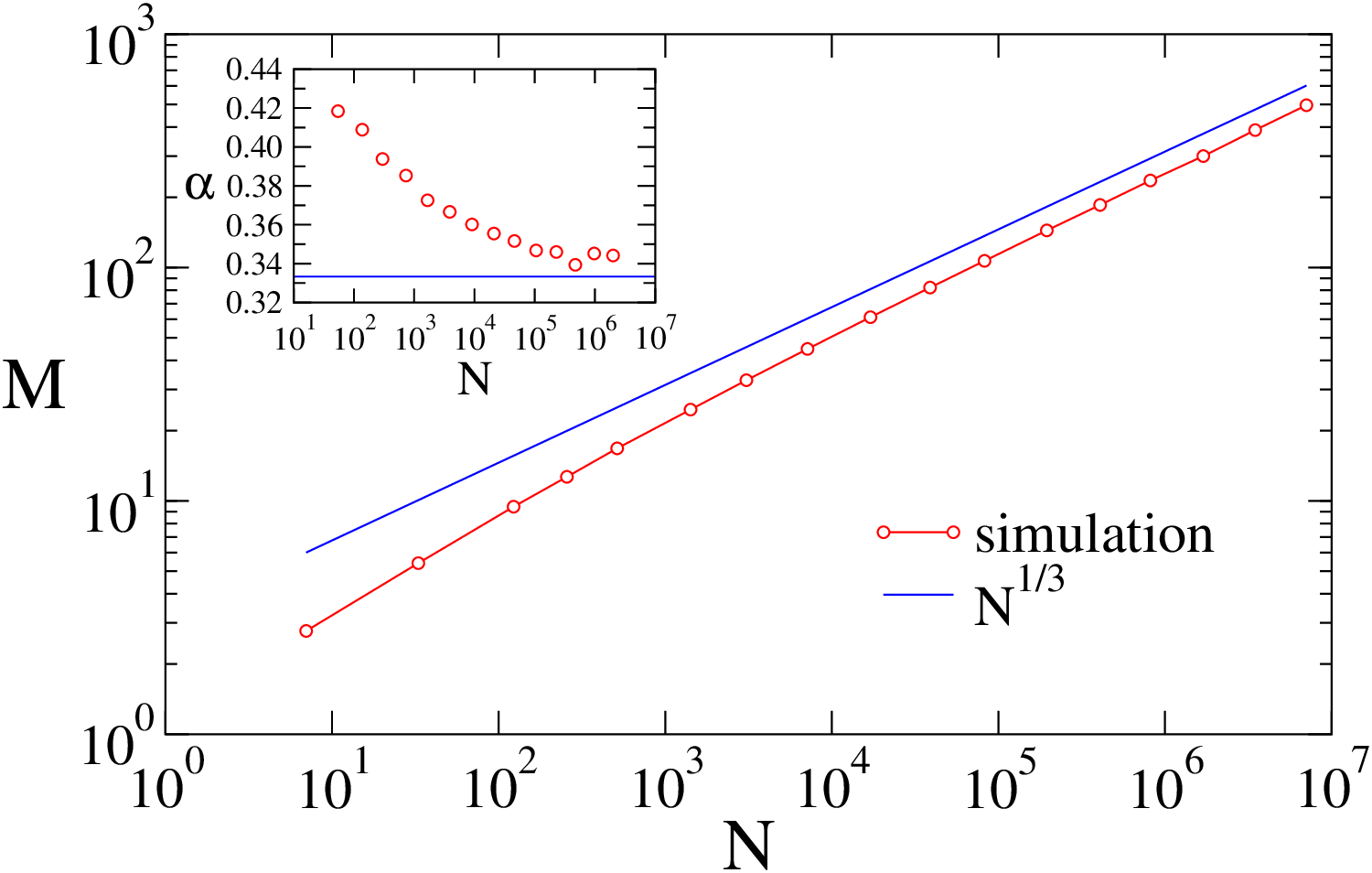}
\caption{The average number of surviving particles $M$ versus the
  initial number of particles $N$. Initially, the particles occupy a
  ball on a three-dimensional lattice. The quantity $M$ is measured
  by fitting the late-time behavior of $n(t)$ to the linear function 
  $a+b\,t^{-1/2}$, as follows from \eqref{regimes}. The inset
  shows the exponent $\alpha \equiv d\ln M/d\ln N$ versus $N$; the
  reference line $\alpha=1/3$ is drawn as well.}
\label{fig-mn}
\end{figure}

We focus on the interesting case $d>2$, and we study the average
number of surviving particles $M$ as a function of the number of
starting particles $N$. Our main result is the scaling law (see
Fig.~\ref{fig-mn})
\begin{equation}
\label{M}
M\sim N^\alpha \qquad {\rm with}\qquad \alpha=\frac{d-2}{d}, 
\end{equation}
which applies for $N\gg 1$. To derive \eqref{M}, we study the time
evolution of the number of particles. We find two time scales: the
diffusion time, $T\sim N^{2/d}$, and the escape time, 
$\tau\sim N^{4/d}$. The vast majority of annihilation events occur on the
diffusion time scale, while no annihilation events events occur beyond
the escape time scale.

The single-species annihilation process \eqref{process} can be
realized either in continuous or in discrete space. In the
continuous-space realization, particles have a finite size and undergo
Brownian motion with diffusion coefficient $D$. In the discrete-space
realization, particles reside in a $d$-dimensional hyper-cubic
lattice.  Each particle occupies a single site and hops with rate $D$
to a neighboring site, chosen at random.  When a particle hops onto an
already occupied site, both particles disappear instantaneously. In
the numerical simulations, we implemented the discrete-space version.

Our focus is the behavior of a finite number of particles. Hence, we
consider initial conditions where a compact region of space is
occupied by a finite number of particles, $N$, while its outer region
is empty. We make two assumptions on the initial arrangement of the
particles: (i) the occupied region is compact, and (ii) the particle
concentration is uniform.  Hence, the number of particles is
proportional to the volume of the region: $N\sim L^d$, where $L$ is
the linear dimension of that domain.  In the simulations, we
implemented a uniform distribution inside a $d$-dimensional ball: all
lattice sites within distance $R$ of the origin are occupied.

The concentration $c({\bf r},t)$ obeys the standard reaction-diffusion
equation
\begin{equation}
\label{ct-eq}
\frac{\partial c}{\partial t} = D\,\nabla^2 c - K\,c^2 \,.
\end{equation}
In writing this equation, we assume that the particles are perfectly
mixed, that is, particle positions are not correlated. Consequently,
the reaction term is quadratic in the concentration. For
single-species annihilation, the reaction equation \eqref{ct-eq} is
valid in dimension $d>2$ \cite{mvs}.  The reaction rate $K$ scales
with the diffusion coefficient $D$ and the particle size $a$ according
to $K\sim D a^{d-2}$ \cite{otb,krb}. For the discrete-space
realization, we identify $D$ with the hopping rate, and $a$ with the
lattice spacing. Without loss of generality, we set $D=1$, $a=1$, and
$K=1$.

Using the reaction-diffusion equation \eqref{ct-eq} and a key
simplifying assumption about the spatial arrangement of the particles,
we can obtain a closed rate equation for the total number of particles
at time $t$, $n(t)=\int d{\bf r}\,c({\bf r}, t)$. First, we integrate
the reaction-diffusion equation \eqref{ct-eq} over the entire space and obtain
\begin{equation}
\label{nt-eq0}
\frac{dn}{dt} = -\int d{\bf r}\,c^2.
\end{equation}
This equation states that the rate of decline in the number of particles
equals the total reaction rate.

In the initial state, particles are confined to a region of space with
a {\em finite} volume. Importantly, the same remains true at all
times.  We expect that the particle ``cloud'' does expand with time,
but nevertheless, the size of this cloud remains finite because the
number of particles is finite. Let us consider a cloud of particles,
confined within a region of volume $V$. In our heuristic analysis, we
assume that the particles remain uniformly distributed inside this
region
\begin{equation}
\label{ct}
c({\bf r}, t) = \frac{n}{V}\,,
\end{equation}
while the concentration vanishes outside this domain, $c({\bf r}, t) =
0$.  By substituting the uniform concentration \eqref{ct} into the
rate equation \eqref{nt-eq0}, we arrive at the closed rate equation
for the average number of particles,
\begin{equation}
\label{nt-eq}
\frac{dn}{dt} = - \frac{n^2}{V}.
\end{equation}
We reiterate that two assumptions were used to derive \eqref{nt-eq},
viz.  (i) the particles are confined to a finite region with volume
$V$, and (ii) the particles are uniformly distributed inside this
volume. But we made no assumptions about the shape of the confining
region.

As the particles diffuse, the volume of the region confining them grows
with time, $V\equiv V(t)$.  Initially, particles are uniformly
distributed and hence $V(t=0)\simeq N$.  Particles that survive the
annihilation process eventually manage to diffuse outside the
initially occupied domain (see figure \ref{fig-ill}). From the
diffusion length scale $\ell\sim t^{1/2}$, we deduce the growth
$V(t)\sim \ell^d\sim t^{d/2}$.  Therefore the confining volume
exhibits two regimes of behavior,
\begin{equation}
\label{Vt}
V(t) \sim 
\begin{cases}
N           & \quad t \ll T,\\
t^{d/2}      & \quad t \gg T.
\end{cases}
\end{equation}
The crossover time scale $T$ can be obtained by matching the two
quantities
\begin{equation}
\label{T}
T \sim N^{2/d}.
\end{equation}
We term $T$ the diffusion time, as this scale characterizes the time
it takes a particle to diffuse outside the initially occupied domain.
As we show below, this time scale separates two regimes of behavior:
an initial regime during which most annihilation events occur, and a
late regime during which the remaining few reaction events occur.

\begin{figure}[t]
\includegraphics[width=0.5\textwidth]{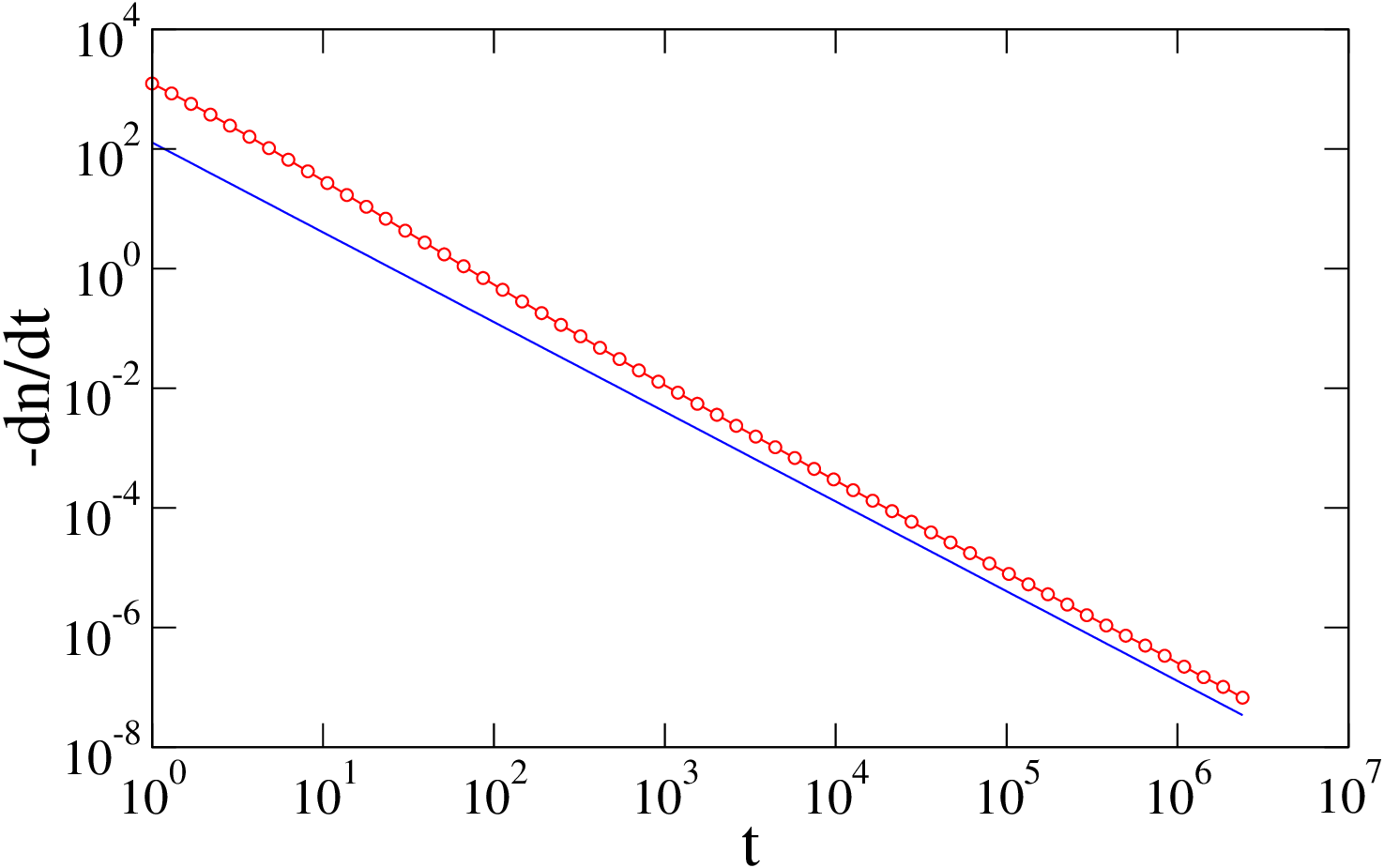}
\caption{The total reaction rate $-dn/dt$ versus time $t$ for a system
  with $N=7153$ particles in three dimensions. Also shown for
  reference is a line with slope $-3/2$.}
\label{fig-dndt}
\end{figure}

In the early regime, the reactions occur within the initially occupied
region with volume $V\sim N$. Equation \eqref{nt-eq} becomes $dn/dt
\sim - n^2/N$, and therefore,
\begin{equation}
\label{nt1}
n(t)\sim N\, t^{-1}
\end{equation}
for $1\ll t\ll T$.  This rapid decay holds as long as most particles
remain within the initially occupied domain.

We estimate the average number of particles that
survive the first phase of the annihilation process by substituting
the time scale \eqref{T} into Eq.~\eqref{nt1} to yield 
\begin{equation}
n(T)\sim N^{\alpha}\, 
\end{equation}
with $\alpha=(d-2)/d$.  Since $n(T)/N\sim N^{-2/d}$, the vast majority
of particles are annihilated at times $t\ll T$.  As we show below, a
finite fraction of particles that survive till time $t\sim T$,
survive ad infinitum, \hbox{$n(\infty)\sim n(T)$}.

In the late regime, the confining volume grows according to 
$V\sim t^{d/2}$, and Eq.~\eqref{nt-eq} becomes $dn/dt \sim - n^2/t^{d/2}$. 
By integrating this equation from $t_0$ till time $t$ we find
\begin{equation}
\label{nt2}
\frac{1}{n(t)}-\frac{1}{n(t_0)}=\frac{1}{t_0^{d/2-1}}-\frac{1}{t^{d/2-1}}\,,
\end{equation}
for $t>t_0\gg T$.  In writing \eqref{nt2} we ignored numerical factors
of order unity.  Setting $t_0\approx T$ and taking the limit
$t\to\infty$ in Eq.~\eqref{nt2}, we arrive at our main result: the
scaling law \eqref{M} for the average number of surviving particles,
$M\equiv n(\infty)$, as a function of the total number of starting
particles $N$. Using numerical simulations we estimate $\alpha=0.34\pm
0.02$ for the exponent governing the scaling law \eqref{M} in three
dimensions (Fig.~\ref{fig-mn}).  As stated above, the reduction in the
number of particles in the second stage of the reaction process is
moderate, $n(\infty)\sim n(T)$, and a finite fraction of particles
that survive past the diffusion time scale, escape annihilation. The
exponent $\alpha$ in \eqref{M} vanishes in the limit $d\to 2$,
consistent with the fact that no particles survive when $d\leq
2$. Also, a finite fraction of the particles survive, $M\sim N$, in
the limit $d\to\infty$.

In summary, the total number of particles exhibits two regimes of
behavior
\begin{equation}
\label{regimes}
\frac{n(t)}{M}\sim 
\begin{cases}
(t/T)^{-1} & 1\ll t\ll T ;\\
1+{\rm const.}\times (t/T)^{(2-d)/2} & T\ll t .
\end{cases}
\end{equation}
Our simulation results confirm the asymptotic behavior $dn/dt\sim
t^{-d/2}$ when $t\gg T$ (see Fig.~\ref{fig-dndt}).  We have used the
asymptotic \hbox{$n(t)\simeq n(\infty)+{\rm const}\times t^{(2-d)/2}$}
to estimate the final number of particles $M\equiv n(\infty)$.  For a
given  $N$, we measure the late time
behavior and fit the number of remaining particles $n(t)$ to a linear
function of $t^{(2-d)/2}$. The intercept of this line yields $M$.

\begin{figure}[t]
\includegraphics[width=0.5\textwidth]{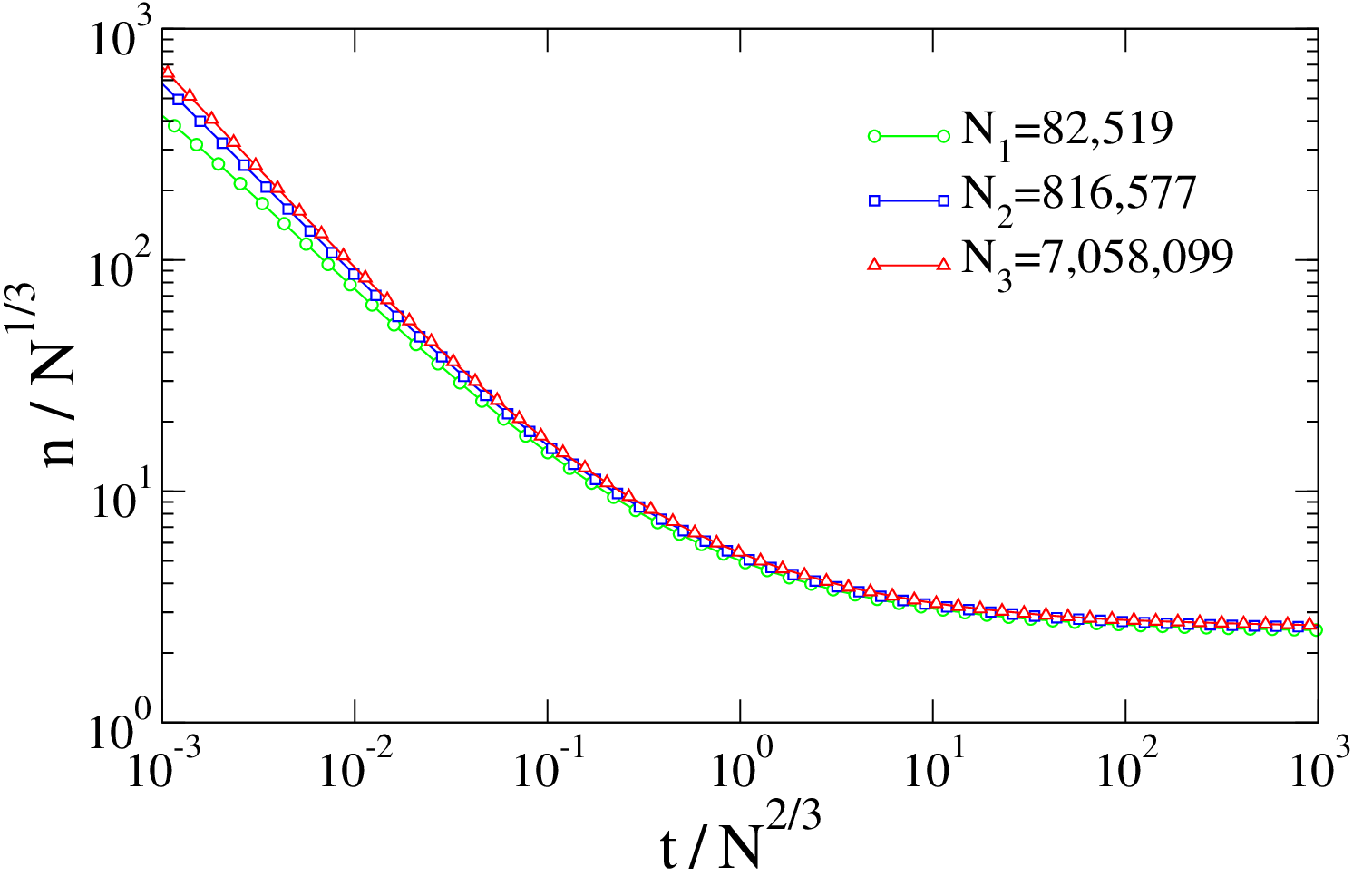}
\caption{The scaled number of particles $n(t)/N^{1/3}$ versus the
  scaled time $t/N^{2/3}$, for three different initial conditions.
  The particle numbers correspond to balls with radii $R_1=27$,
  \hbox{$R_2=58$}, and $R_3=119$. The simulation results represent an average
  over $10^3$, $10^2$, and $10^1$ independent realizations.}
\label{fig-nt}
\end{figure}

A useful way to express the time-dependent behavior \eqref{regimes} is
through the finite-size scaling form \cite{fb,bl}
\begin{equation}
\label{scaling} 
n(t)\simeq N^{\alpha}\, F\left(t/N^{2/d}\right).
\end{equation}
This form reflects that, in properly scaled units, the statistics 
become independent of the number of particles in the large-$N$ limit
(see Fig.~\ref{fig-nt}).  The scaling function has two limiting
behaviors
\begin{equation}
F(x) \sim 
\begin{cases}
x^{-1} & x\ll 1; \\
1+ {\rm const.}\times x^{(2-d)/2} & x\gg 1.
\end{cases}
\end{equation}

Since the number of particles is finite, the time at which the final
reaction event takes place is also finite. The final reaction
reduces the number of particles by two, and hence, the time of the
final reaction event $\tau$ can be estimated from
\hbox{$n(\tau)-n(\infty)=2$}.  Rewriting 
\eqref{regimes} as \hbox{$n(t)-n(\infty)\sim
  N^{2\alpha}\,t^{(2-d)/2}$} we obtain the ``escape'' time scale
\begin{equation}
\label{tau}
\tau\sim N^{4/d}\,.
\end{equation}
This time scale also sets a length scale $\rho \sim \tau^{1/2}\sim
N^{2/d}$ for the escape process. Particles that manage to diffuse a
distance comparable to the escape length scale $\rho$ survive
forever. Note that the escape length scale grows quadratically with
the linear dimension of the initial domain $\rho\sim L^2$.

To summarize, the diffusion time scale $T\sim N^{2/d}$ is the time it
takes particles to diffuse outside the initially occupied domain, and
nearly all annihilation events occur in this time window.  The escape
time $\tau\sim N^{4/d}$ is the time at which the reaction process
stops. The universal relationship, $\tau\sim T^2$, connects these time
scales.

We now estimate the average lifetime of an annihilated
particle. Nearly all particles disappear in the first time regime,
$t\ll T$. From the density decay \eqref{nt1}, we get $t^{-2}$ for the
annihilation rate and therefore the average lifetime is $\langle
t\rangle \sim \int^T dt\,t\,t^{-2}$. Using $T\sim N^{2/d}$, we obtain
\begin{equation}
\label{tav}
\langle t\rangle \sim \ln N. 
\end{equation}
Hence, the lifetime of reacting particles is relatively short, growing
only logarithmically with system size. This feature allows us to
simulate a large number of particles.  Our Monte Carlo simulations
utilize ${\cal O}(N)$ memory but require a quadratic number of
operations per surviving particle, per unit time. Thanks to
\eqref{tav}, the overall complexity of our {\it
  brute-force} simulations is only ${\cal O}(N^2\ln N)$.

We now  mention several extensions of the above results. First,
we consider the multi-particle annihilation process
\begin{equation}
\label{multi}
\underbrace{A+A+\cdots+A}_{m}\to \emptyset, 
\end{equation}
which generalizes the binary reaction process \eqref{process} to an
arbitrary number $m$ of reacting particles. The basic rate equation
\eqref{nt-eq} becomes $dn/dt=-n^m/V$. By generalizing the above
analysis, we obtain
\begin{equation}
M\sim N^{(d-d_c)/d},\qquad d_c=\tfrac{2}{m-1}.
\end{equation}
This scaling behavior holds above the critical dimension, $d>d_c$; for
$d\leq d_c$, all particles disappear. The diffusion time remains
$T\sim N^{2/d}$, but the escape time is $m$-dependent, $\tau \sim
N^{(2m)/[d(m-1)]}$.

Thus far, we have implicitly assumed that the dimension of the
initially-occupied region $\delta$ equals the spatial dimension
$d$. We now consider the situation where $\delta<d$. For example, when
particles initially occupy a two-dimensional disk in three dimensions
then $\delta=2$ and $d=3$. In general, the number of particles scales
with the linear dimension as follows $N\sim L^\delta$. To address this
problem, we use an alternative, probabilistic, approach.

We demonstrate this approach for the case $d=\delta$. Initially, the
particles occupy a compact domain of volume $L^d$ and the typical
distance between particles equals $1$.  At later times, the surviving
particles still occupy the same domain of volume $L^d$, but the
typical distance between neighboring particles grows to $\ell\gg 1$.
We take a test particle, located in the bulk of the domain, and
estimate its survival probability, assuming that all other particles
survive.  Around the test particle, we draw spheres of radius
$n\ell$, with $n=1,2,\ldots,L/\ell$ and $n^{d-1}$ particles on each
spherical shell.  For two Brownian particles separated by distance
$r$, the probability that they never meet is
$1-r^{-(d-2)}$\cite{sr}. The product of such probabilities,
\begin{equation}
\label{product}
\prod_{\ell=1}^{L/\ell} \left[1-\frac{1}{(n\ell)^{d-2}}\right]^{n^{d-1}},
\end{equation}
gives a lower bound for the survival probability. If this product if
finite in the limit $N\to\infty$, then the survival probability is
finite.  The product is finite if and only if its logarithm is finite
and hence,
\begin{equation}
\label{log}
\frac{1}{\ell^{d-2}}\sum_{\ell=1}^{L/\ell} n \sim \frac{L^2}{\ell^d}\sim 1\,. 
\end{equation}
Therefore, to guarantee that the survival probability is finite we
must choose $\ell^d\sim L^2$, and using the number of surviving
particles, \hbox{$M\sim (L/\ell)^d$}, we recover \eqref{M}.

This probabilistic argument can be generalized to situations where the
initially-occupied region has dimension $\delta$. Replacing the power
$n^{d-1}$ in the product \eqref{product} with $n^{\delta-1}$, and
repeating the steps above, yields the average number of surviving
particles
\begin{equation}
\label{general}
M \sim 
\begin{cases}
N^{(d-2)/\delta}             & d-\delta<2,\\
N(\ln N)^{-1}              & d-\delta=2,\\
N                         & d-\delta>2.
\end{cases}
\end{equation}
For example, for a two-dimensional disk in three dimensions we have
$M\sim N^{1/2}$.  In general, the co-dimension $d-\delta$ governs the
behavior. A finite fraction of the particles eternally survive when the
co-dimension is sufficiently large, $d-\delta>2$, while the number of
surviving particles grows algebraically with $N$ when the co-dimension
is small, $d-\delta<2$.

In summary, we studied reaction kinetics of single-species
annihilation starting with a finite number of particle. Using the rate
equation approach and heuristic arguments, we derived finite-size
scaling properties of the time-dependent number of particles. Systems
with a finite, yet large, number of particles exhibit universal
behavior, once time and particle number are properly scaled. We have
shown that when $d>2$, a finite number of particles escape the
reaction process. This number scales sub-linearly with the total
number of particles. In addition to the diffusion time that
characterizes the process, we also found a much larger time scale
which characterizes the escape process.

We stress that our initial conditions: a ``droplet'' containing a
finite number of reactants has physically relevance. While we focused
on spherical droplets in our simulations, we expect the same
qualitative behaviors occurs for other compact initial conditions, say
for particles occupying a $d-$dimensional cube and even for
non-compact initial conditions as long as all moments of the
distribution function characterizing the distance to the origin are
finite.

Our analysis provides scaling predictions for average quantities such
the average number of surviving particles. It would be interesting to
investigate the distribution of the number of surviving
particles. Given the finite-size scaling form \eqref{scaling}, we
expect that a universal distribution emerges when the number of
particle is large. The shape of this distribution is an interesting
topic for future studies. Of special interest is the probability
$E_d(N)$ that all particles disappear. Using the probabilistic
approach \eqref{product}--\eqref{log}, we obtained the following
estimate for this extinction probability: $\ln E_d(N) \sim -
N^\alpha\,\ln N$.

The behavior of other reaction processes is another topic for further
research. We expect that our scaling results hold for the
closely-related coalescence process, $A+A\to A$. However, the
aggregation process, $A_i+A_j\to A_{i+j}$, with mass-dependent
diffusion coefficients, e.g. $D_k\sim k^{-\mu}$, appears to be more
challenging.  Another related problem is two-species annihilation,
$A+B\to \emptyset$, where the critical dimension is $d_c=4$. Starting
with equal numbers of $A$ and $B$ particles, one gets 
$n\sim N\,t^{-d/4}$ when $d<4$ and $n\sim N\,t^{-1}$ otherwise
\cite{zo,tw}. The behavior above the critical dimension should
coincide with \eqref{M}, but the behavior below the critical dimension
is intriguing. Recurrence properties of Brownian particles again imply
that no particle eternally survive when $d\leq 2$.  Using the
diffusion time \eqref{T}, repeating the scaling arguments above yields
$M\sim N^{1/2}$ for all $2<d<4$. It would be interesting to
investigate this problem using theoretical or computational methods.

\medskip\noindent 

We are indebted to Nadav Shnerb for useful discussions and we
acknowledge support from US-DOE grant DE-AC52-06NA25396 (EB).


\begin{thebibliography}{99}
\bibitem{odd} 
      When the initial number of particles is odd, a single
      particle survives when $d\leq 2$.
\end{thebibliography}
\end{document}